%% file: paper_1.tex
\documentclass[twocolumn,showpacs,aps,prl,superscriptaddress]{revtex4}

\usepackage{graphicx}
\usepackage{dcolumn}
\usepackage{amsmath}
\usepackage{epsfig}

\input pubboard/babarsym

\newcommand{\z}{\ensuremath{{\mathsf z}}\xspace}
\newcommand{\Imz}{\ensuremath{\rm Im\, \z}}
\newcommand{\ImzZero}{\ensuremath{\rm Im\, \z_0}}
\newcommand{\ImzOne}{\ensuremath{\rm Im\, \z_1}}
\newcommand{\Rez}{\ensuremath{\rm Re\, \z}}

\newcommand{\dG}{\ensuremath{ \Delta \Gamma }}
\newcommand{\absqop}{\ensuremath{|q/p|}}
\newcommand{\RezDG}{\ensuremath{\Rez}\,\dG}
\newcommand{\RezZeroDG}{\ensuremath{\Rez_{0}\,\dG}}
\newcommand{\RezOneDG}{\ensuremath{\Rez_{1}\,\dG}}

\def\dt {\ensuremath{\Delta t }}
\def\dm {\ensuremath{\Delta m }}

\newcommand{\BABARPubYear}    {07}
\newcommand{\BABARPubNumber}  {063}

\newcommand{\SLACPubNumber} {13010}

\def\figurebox#1#2#3{%
    \def\arg{#3}%
    \ifx\arg\empty
    {\hfill\vbox{\hsize#2\hrule\hbox to #2{\vrule\hfill\vbox to #1{\hsize#2\vfill}\vrule}\hrule}\hfill}%
    \else
    {\hfill\epsfbox{#3}\hfill}%
    \fi}

\begin{document}

\preprint{\babar-PUB-\BABARPubYear/\BABARPubNumber} 
\preprint{SLAC-PUB-\SLACPubNumber} 

\begin{flushleft}
\babar-PUB-\BABARPubYear/\BABARPubNumber\\
SLAC-PUB-\SLACPubNumber\\
\end{flushleft}

\title{
{\large \bf \boldmath Search for \CPT and Lorentz Violation in \Bz-\Bzb\ Oscillations with Dilepton Events} 
}

\input pubboard/authors_aug2007

\begin{abstract}
We report results of a search for \CPT and Lorentz violation in \Bz -\Bzb
oscillations using inclusive dilepton events from 232 million \upsbb\ decays 
recorded by the \babar\ detector at the \pep2\ \BF\ at SLAC.  
We find $2.8\sigma$ significance, compatible with no signal, for variations in the 
complex \CPT violation parameter \z\ at the Earth's sidereal frequency and extract 
values for the quantities $\Delta a_\mu$ in the general Lorentz-violating 
standard-model extension. The spectral powers for variations in \z 
over the frequency range 0.26\,year$^{-1}$ to 2.1\,day$^{-1}$ are also 
compatible with no signal.
\end{abstract}

\pacs{13.25.Hw, 12.15.Hh, 11.30.Er}

\maketitle
It was shown recently~\cite{Greenberg_2002} that an interacting quantum field 
theory need not be local for \CPT violation to imply  
violation of Lorentz invariance.
In the general Lorentz-violating standard-model extension
(SME)~\cite{Kostelecky_1}, the parameter for \CPT violation in 
neutral meson oscillations depends on the 4-velocity of the 
meson~\cite{Kostelecky_2}.

We report a search for this effect using \upsbb\ decays recorded by the 
\babar\ detector at the PEP-II asymmetric-energy $e^+e^-$ collider.
Any observed \CPT violation should vary with a period of one sidereal day
($\simeq 0.99727$\,solar days) as the \FourS boost direction follows the 
Earth's rotation with respect to the distant stars~\cite{Kostelecky_3}.

The physical states of the \Bz-\Bzb\ system are 
\begin{equation}
\begin{array}{rcl}
|B_L\rangle&=&p\sqrt{1-\z}|\Bz\rangle +q\sqrt{1+\z}|\Bzb\rangle, \\ [0.1cm]
|B_H\rangle&=&p\sqrt{1+\z}|\Bz\rangle -q\sqrt{1-\z}|\Bzb\rangle,
\end{array}
\label{eq:mass_eigenstates_cpt}
\end{equation}
where $L$ ($H$) labels the ``light'' (``heavy'') eigenstate of the 
effective Hamiltonian.  The complex parameter \z vanishes 
if \CPT is conserved; \T invariance implies $\absqop=1$.

In the SME, \CPT- and Lorentz-violating coupling coefficients $a_\mu^{q_i}$ 
for the two valence quarks in the \Bz\ meson are contained in quantities 
$\Delta a_\mu = r_{q_1}a_\mu^{q_1} - r_{q_2}a_\mu^{q_2}$, where the $r_{q_i}$ 
are due to quark-binding and normalization effects.
The \CPT parameter \z depends on the meson 4-velocity 
$\beta^\mu=\gamma(1,\vec{\beta})$ in each experiment's observer frame as~\cite{Kostelecky_2}
\begin{equation}
\z \simeq \frac{\beta^\mu \Delta a_\mu}{\Delta m - i\Delta\Gamma/2},
\label{eq:zBDa}
\end{equation}
where $\beta^\mu \Delta a_\mu$ is real and varies with sidereal time due to 
the rotation of $\vec{\beta}$ relative to the constant vector $\Delta\vec{a}$. 
The magnitude of the decay rate difference 
$\Delta\Gamma \equiv \Gamma_H - \Gamma_L$ is known to be small compared
to the \Bz-\Bzb\ oscillation frequency $\dm \equiv m_H - m_L$; hence 
Eq.~\ref{eq:zBDa} constrains
\begin{equation}
\dm{\Rez} \simeq 2\dm(\dm/\dG){\Imz} \simeq \beta^\mu \Delta a_\mu.
\label{eq:BDaRezImz}
\end{equation}

Limits on analogous flavor-dependent $\Delta a_\mu$ specific to \KzKzb\ 
oscillations~\cite{KTEV1} and to \DzDzb\ oscillations~\cite{FOCUS} have been 
reported by the KTeV and FOCUS collaborations, respectively. 
KTeV has also reported a limit on sidereal variation of the phase $\phi_{+-}$ 
of the \CP-violating amplitude ratio $\eta_{+-} = {\cal A}(K_L \to \pi^+\pi^-)/{\cal A}(K_S \to \pi^+\pi^-)$~\cite{KTEV2}.

We adopt the basis $(\hat{x},\hat{y},\hat{z})$ for the 
rotating laboratory frame and the basis $(\hat{X},\hat{Y},\hat{Z})$ for the Sun-centered 
non-rotating frame containing $\Delta\vec{a}$~\cite{Kostelecky_4}.
$\hat{Z}$ is parallel to the Earth's rotation axis, $\hat{X} (\hat{Y})$ 
is at right ascension $0^\circ (90^\circ)$, and $\hat{y}$ is at declination 
$0^\circ$. We take $\beta^\mu$ for each $B$ meson to be the \FourS 
4-velocity, and choose $\hat{z}$ to lie along $-\vec{\beta}$. 
The event sidereal time $\hat{t}$ is given by the right ascension of $\hat{z}$ 
as it precesses around $\hat{Z}$ at the sidereal frequency 
$\Omega = 2\pi$\,rad/sidereal-day.
We find $\hat{t} = 14.0$ sidereal-hours at the Unix epoch (00:00:00 UTC, 
1 Jan.\ 1970) from the latitude ($37.4^\circ$\,N) and longitude 
($122.2^\circ$\,W) of \babar\ and the \FourS boost 
($\langle\beta\gamma\rangle \simeq 0.55$ toward $37.8^\circ$ east of south), 
which also yield $\cos\chi = \hat{z}\cdot\hat{Z} = 0.628$ in Eq.~\ref{eq:BDa}:
\begin{equation}
\begin{array}{rcl}
\hspace*{-0.1cm}\beta^\mu  \Delta a_\mu& = & \gamma\left[\Delta a_0 - \beta\Delta a_Z\cos\chi \right . \\
                       &   & \left . -\beta\sin\chi\left(\Delta a_Y\sin\Omega\hat{t} + \Delta a_X\cos\Omega\hat{t}\,\right)\right]. 
\end{array}
\label{eq:BDa}
\end{equation}

Neutral $B$ mesons from \FourS decay evolve in orthogonal flavor states 
until one decays, after which the flavor of the other continues to oscillate. 
We use {\it direct} semileptonic decays ($b \to X\ell\nu$, where $\ell = e$ 
or $\mu$) to tag the flavor of each $\Bz(\Bzb)$ by the charge of the lepton 
$\ellp(\ellm)$. The decay rate for opposite-sign dilepton ($\ellp\ellm$) 
events is
\begin{equation}
\begin{array}{rcl}
\hspace*{-0.45cm}N^{+-}\!\!\!&\propto&\!e^{-|\dt|/\tau_{\Bz}} \left\{(1+|{\z}|^2)\cosh(\dG \dt /2) \right . \\
           &       & \hspace{1.4cm} +\,(1-|{\z}|^2) \cos(\dm \dt) \\ 
           &       &\!\!\left . -2\,{\Rez}\sinh(\dG \dt /2) + 2\,{\Imz}\sin(\dm \dt)\right\}\!.
\label{eq:decayrate}
\end{array}
\end{equation}
We define $1/\tau_{\Bz}$ to be the average neutral $B$ decay rate, and 
$\dt\equiv t^+ - t^-$, where $t^+(t^-)$ is the proper time for one of a pair of $B$ mesons 
to decay to $\ellp(\ellm)$. We make the approximation $\sinh(\dG\dt/2) \simeq \dG\dt/2$, 
which is valid for the range $|\dt| < 15$\,ps used in this analysis.
We use $|\dG|= 6 \times 10^{-3}\ps^{-1}$ in the $\cosh(\dG \dt /2)$ term, consistent with 
the value reported in Ref.~\cite{BaBarCPT}.

The asymmetry between the decay rates at $\dt>0$ and $\dt<0$ compares the probabilities $P(\Bz \to \Bz)$ and 
$P(\Bzb \to \Bzb)$. Omitting second-order terms in \z gives
\begin{equation}
A_{CPT}(\dt) \simeq \frac{-{\Rez}\,\dG \dt + 2\,{\Imz}\sin(\dm\dt)}{\cosh(\dG \dt /2) + \cos(\dm\dt)}. 
\label{eq:acpt1}
\end{equation}

The \babar\ detector is described elsewhere~\cite{ref:babar}. 
We use about 232 million \upsbb decays and 16\,\invfb of off-resonance 
data, from 40\,\mev below the \FourS resonance, collected during 1999--2004 
to search for variations in \z with sidereal time of the form
\begin{equation}
\z = \z_0 + \z_1\cos{(\Omega\hat{t} + \phi)}.
\label{eq:z_form}
\end{equation}
For long data-taking periods, any day/night variations in detector response 
tend to cancel over sidereal time.

We have previously measured~\cite{thePRL} time-integrated values of \Imz\ and \RezDG\ 
from the \dt\ distribution of the same events. 
Here, we measure \ImzZero, \RezZeroDG, \ImzOne, and 
\RezOneDG\ by extending the likelihood fit to include the event sidereal time $\hat{t}$, and
extract values for the SME quantities $\Delta a_\mu$. In a complementary 
approach, we also measure the spectral power of periodic variations in \z over a wide 
frequency band using the periodogram method~\cite{pgram} developed to study 
variable stars.

The event selection is the same as in Ref.~\cite{thePRL}.
Briefly, we suppress non-\BB background by event-shape and event-topology 
requirements, and select events having at least two well-identified lepton 
candidates with momenta 0.8\,--\,2.3\,\gevc in the \FourS rest frame that are 
not part of reconstructed $\jpsi,\psi (2S) \rightarrow e^+e^-, \mu^+\mu^-$ 
decays or photon conversions. Lepton candidates must have at least one 
$z$-coordinate measurement in the silicon vertex tracker to allow \dt\ 
to be well-measured.  
We reject events in which either of the two highest-momentum lepton candidates 
(the {\it dilepton}) is classified as a {\it cascade} lepton from a 
$b \rightarrow (c, \tau) \rightarrow \ell$ transition by a neural-network 
algorithm that uses as input variables the momenta and opening angle of the 
two leptons together with the event's visible energy and missing momentum.
The selected dilepton sample comprises 1.18 million opposite-sign events 
and 0.22 million same-sign events.

We estimate the \FourS decay point in the transverse plane with a $\chi^2$-fit
using the transverse distances to the two lepton tracks and the beam-spot.
To measure $\dt$, we assume each lepton originates from a direct $B$ meson 
decay at the point on the lepton track with the least transverse distance to 
the \FourS. The component $\Delta z$, along the Lorentz boost,
of the distance between these two points yields 
$\dt=\Delta z/ \langle\beta\gamma\rangle c$. For opposite-sign events 
$\Delta z = z^+ - z^-$; for same-sign events we use $|\Delta z|$.

We model the $\dt$-distribution of the dilepton sample with the 
probability density functions (PDFs) used in Ref.~\cite{thePRL} to represent 
contributions from \BzBzb and \BpBm decays and non-\BB events. 
The latter are estimated, using off-resonance data, to be 3.1\% of the sample. 
The fit to data determines that 59\% of the \BB events are \BpBm decays. 
With minor \BB background contributions fixed to values from Monte Carlo 
(MC) simulation, the fit to data also determines the fractions of \BzBzb and 
\BpBm decays that are {\it signal} events ($\simeq 80$\%) with two direct 
leptons, and the fractions ($\simeq 10$\%) that are events with one direct 
lepton and a $b \rightarrow c \rightarrow \ell$ cascade decay of the other $B$ 
meson. Same-sign dilepton events are retained primarily to improve 
the determination of these fractions.

Each PDF is a convolution of a decay rate in $\dt$ with a resolution function 
that is a sum of Gaussians or, for events with a cascade lepton, its 
convolution with one or two double-sided exponentials accounting for the 
lifetimes of intermediate $\tau$ or $D_{(s)}$ meson states, respectively.
We use a sum of three Gaussians for signal events. The fit to data determines 
their fractions and also their widths except that of the widest, which is 
fixed to 8\,ps.
For leptons from different $B$ mesons, our \BzBzb decay rate  
contains \z to first-order (cf.\ Eq.~\ref{eq:decayrate}) for opposite-sign 
events and is 
$\propto e^{-|\dt|/\tau_{\Bz}}\left\{\cosh(\dG \dt /2) - \cos(\dm \dt)\right\}$
for same-sign events; for \BpBm\ decays, it is 
$\propto e^{-|\dt|/\tau_{\Bpm}}$. For leptons from the same $B$ meson, the 
decay rates are exponentials with effective lifetimes determined from MC 
simulation. Dilution factors are included to account for wrong flavor tags in 
cascade decays.

Each event's timestamp yields the time elapsed since the Unix epoch. 
We use this time, folded over one sidereal day and shifted in phase by 
$14.0$ sidereal-hours, for $\hat{t}$.

We extract \z from a two-dimensional maximum likelihood fit to the 
opposite-sign and same-sign data events binned separately in $\dt$ and 
$\hat{t}$. The likelihood function in $\dt$ for each of the 24 sidereal-time 
slices contains a common sum of the PDFs, and \z varies with $\hat{t}$ as in Eq.~\ref{eq:z_form}.
The likelihood fit corresponds to $A_{CPT}$ in Eq.~\ref{eq:acpt1}. We obtain  
the values for \z\ and $\phi$ reported in Table~\ref{tab:Syst} (upper left).
The statistical correlation between \ImzZero\ and \RezZeroDG\ is 76\%; 
between \ImzOne\ and \RezOneDG\ it is 79\%.  

\begin{table*} [!htb]
\caption{Asymmetry parameter values, with statistical errors, for $A_{\CPT}$ in 
         Eq.~\ref{eq:acpt1} (upper left) and with SME constraint in Eq.~\ref{eq:acpt2} 
         (upper right). Equation~\ref{eq:z_form} implies $\z_1 \to -\z_1$ for $\phi \to \phi +\pi$. 
         Systematic uncertainties are shown in lower part of Table.}
\begin{center}
\begin{tabular}{lccccccccc}
\hline \hline \\ [-0.3cm]
                                        & \multicolumn{5}{c}{Without SME constraint}                                                                         & & \multicolumn{3}{c}{With SME constraint}              \\ 
\cline{2-6} \cline{8-10} \\ [-0.3cm]
{\bf \boldmath $A_{\CPT}$ parameter}    &  $\ImzZero$        &  $\RezZeroDG$              &  $\ImzOne$         &  $\RezOneDG$               &  $\phi$        & & $\ImzZero$         & $\ImzOne$          &  $\phi$        \\
                                        & $(\times 10^{-3})$ & $(\times 10^{-3}\ps^{-1})$ & $(\times 10^{-3})$ & $(\times 10^{-3}\ps^{-1})$ &  (rad)         & & $(\times 10^{-3})$ & $(\times 10^{-3})$ &  (rad)         \\
\cline{2-6} \cline{8-10} \\ [-0.3cm]
{\bf Value from fit}                    & $-14.2\pm 7.3$     & $-7.3\pm 4.1$              & $-24\pm 11$        & $-18.5\pm 5.6$             & $2.63\pm 0.31$ & & $-5.2\pm 3.6$      & $-17.0\pm 5.8$     & $2.56\pm 0.36$ \\ [0.1cm]
{\bf Systematic effects}                &                    &                            &                    &                            &                & &                    &                    &                \\ 
$\tau_{\Bz}$, $\tau_{\Bpm}$, $\dm$, \dG &  $\pm0.7$          &  $\pm0.4$                  &  $\pm0.6$          &  $\pm0.5$                  & $\pm0.05$      & & $\pm0.4$           & $\pm0.7$           & $\pm0.01$      \\
SVT alignment, $z$ scale                &  $\pm0.6$          &  $\pm1.5$                  &  $\pm2.0$          &  $\pm1.1$                  & $\pm0.20$      & & $\pm1.7$           & $\pm1.4$           & $\pm0.15$      \\
PDF resolution models                   &  $\pm2.0$          &  $\pm1.0$                  &  $\pm2.5$          &  $\pm1.2$                  & $\pm0.02$      & & $\pm0.8$           & $\pm1.0$           & $\pm0.01$      \\
Background fractions                    &  $\pm0.1$          &  $\pm0.1$                  &  $\pm0.2$          &  $\pm0.2$                  & $\pm0.01$      & & $\pm0.2$           & $\pm0.3$           & $\pm0.01$      \\
Sidereal phase                          &  $\pm0.0$          &  $\pm0.0$                  &  $\pm0.0$          &  $\pm0.0$                  & $\pm0.03$      & & $\pm0.0$           & $\pm0.0$           & $\pm0.03$      \\ [0.1cm]
{\bf Total syst. error}                 &  $\pm2.2$          &  $\pm1.8$                  &  $\pm3.3$          &  $\pm1.7$                  & $\pm0.21$      & & $\pm1.9$           & $\pm1.9$           & $\pm0.15$      \\
\hline
\hline
\end{tabular}\label{tab:Syst}
\end{center}
\end{table*}

Table~\ref{tab:Syst} shows the sources of systematic uncertainties in the 
asymmetry parameters. Separate contributions are added in quadrature in 
the totals. We vary separately $\tau_{\Bz}$, $\tau_{\Bpm}$, and $\dm$ by 
$1\sigma$ from their known values~\cite{PDG06}, and vary $|\dG|$ over the 
range 0\,--\,0.1\,ps$^{-1}$ to allow $3\sigma$ deviations from the value 
reported in Ref.~\cite{BaBarCPT}.
Fixed parameters in the PDF resolution functions for non-signal events are 
varied separately by 10\%, motivated by a comparison of resolution parameters 
fitted to signal events in data and MC simulation.
The fractions of the $D_{(s)}$ meson components in background cascade decays 
are also varied by 10\%. 
The effects of possible internal misalignments of the silicon vertex tracker 
(SVT) and uncertainty in the absolute $z$-scale are evaluated in \BzBzb\ MC samples. 
The clock that sets the event timestamps is governed by the PEP-II 
master oscillator, which is stable to within 0.001\% of its set frequency.
Resynchronization of the clock with U.S.\ time standards at intervals of 
less than four months limits relative sidereal phase errors   
to less than 0.2\%. Another small uncertainty in sidereal phase arises 
in calculating the \FourS boost's right ascension. 
We use $e^+e^-\rightarrow \mu^+\mu^-(\gamma)$ data events, with 
true $\Delta z = 0$, to check for sidereal variations in measured 
$\Delta z$ that could mimic a Lorentz-violation signal.  
The measured amplitude $(0.022\pm 0.025)\,\mu$m and mean  
$(0.030\pm 0.018)\,\mu$m are sources of negligible uncertainties. 
At the solar-day frequency, the amplitude is $(0.028\pm 0.025)\,\mu$m.

\begin{figure}[!htbp]
\begin{center}
\includegraphics[width=8.8cm]{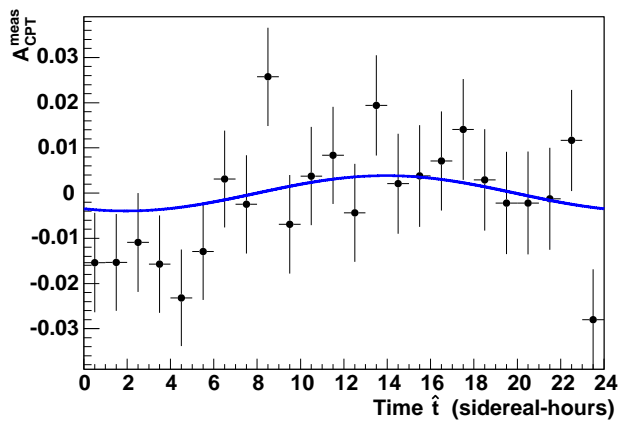}
\caption{Asymmetry $A^{\rm meas}_{CPT}$ for opposite-sign dilepton events 
         with $|\dt| > 3$\,ps versus sidereal time. 
         The sample includes event types, e.g.\ \BpBm\ decays, for which 
         $A_{CPT} = 0$. 
         The curve is a projection, for $|\dt| > 3\ps$, using 
         results of the two-dimensional likelihood fit for $|\dt| < 15\ps$.}
\label{fig:CPTAsyData1}
\end{center}
\end{figure}

\begin{figure}[!htbp]
\begin{center}
\includegraphics[width=8.8cm]{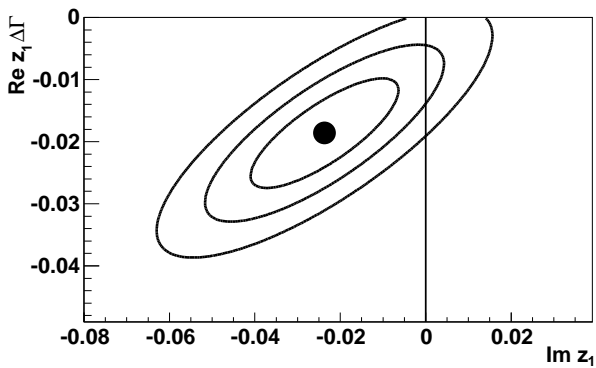}
\caption{Contours indicating $1\sigma$, $2\sigma$, and 
         $3\sigma$ significance, around the central values of \ImzOne\ and 
         \RezOneDG\ (solid circle).}
\label{fig:CPTAsyData2}
\end{center}
\end{figure}

In Fig.~\ref{fig:CPTAsyData1} we plot the sidereal-time dependence of the 
measured asymmetry $A^{\rm meas}_{CPT}$ for the opposite-sign dilepton events
with $|\dt| > 3\ps$, thereby omitting highly-populated bins where 
any asymmetry is predicted to be small.
Figure~\ref{fig:CPTAsyData2} shows confidence level contours for \ImzOne\ and 
\RezOneDG. The significance for sidereal variations in \z, 
characteristic of \CPT and Lorentz violation, is $2.8\sigma$.


The results of the fit described above are 
compatible with the SME constraint 
${\RezDG} \simeq 2\dm\,\Imz$ (Eq.~\ref{eq:BDaRezImz}) for 
$\dm = 0.507$\,ps$^{-1}$~\cite{PDG06}. We repeat the likelihood fit subject to this 
constraint. The asymmetry in Eq.~\ref{eq:acpt1} becomes 
\begin{equation}
A_{CPT}(\dt) \simeq  \frac{2\,{\Imz}\left\{ -\dm\dt + \sin(\dm\dt)\right\}}{\cosh(\dG \dt /2) + \cos(\dm\dt)}.  
\label{eq:acpt2}
\end{equation}
We obtain the results reported in Table I (right).
The statistical correlation between \ImzOne\ and $\phi$ is 48\%. The 
significance for sidereal variations in \z\ is again $2.8\sigma$.
We obtain consistent results for \ImzZero, \ImzOne, and $\phi$ when 
second-order terms (Eq.~\ref{eq:decayrate}) of form 
$|\z|^2 = \rho^2\cos^2(\Omega\hat{t} + \phi)$, motivated 
by finding $|\ImzOne| > |\ImzZero|$, are included in 
the likelihood fit to data with $\rho^2$ as a free parameter.

We use Eqs.~\ref{eq:BDaRezImz}, \ref{eq:BDa}, and \ref{eq:z_form} to extract the SME quantities
\begin{eqnarray*}
\Delta a_0 - 0.30\Delta a_Z         & \simeq & (-3.0  \pm 2.4)(\Delta m/\dG)\times 10^{-15}\,{\rm GeV},\\
\Delta a_X                          & \simeq & (-22 \pm 7)(\Delta m/\dG)\times 10^{-15}\,{\rm GeV},\\
\Delta a_Y                          & \simeq & (-14^{+10}_{-13})(\Delta m/\dG)\times 10^{-15}\,{\rm GeV}.
\end{eqnarray*}
  
We now use the periodogram method~\cite{pgram} to compare the spectral power 
for variations in \z\ at the sidereal frequency with those in a wide band of 
surrounding frequencies. The spectral power at a test frequency $\nu$ is 
\begin{equation}
P(\nu) \equiv \frac{1}{N\sigma^2_w}\Bigl|\sum_{j=1}^{N}w_j e^{2i\pi\nu T_j}\Bigr|^2,
\end{equation}
where the data, comprising $N$ measurements $w_j$ made at times $T_j$, have variance $\sigma^2_w$. 
Here, $T_j$ is the time elapsed since the Unix epoch for 
opposite-sign dilepton event $j$, and the weights 
$w_j = \dm\dt_j - \sin(\dm\dt_j)$ are suited to the study of periodic variations 
in \z\ according to Eq.~\ref{eq:acpt2}. 

In the absence of an oscillatory signal, the probability that $P(\nu)$ exceeds 
a value $S$ at a given frequency is $\exp (-S)$; if $M$ independent 
frequencies are tested, the largest $P(\nu)$ value exceeds $S$ with probability
\begin{equation}
\Pr\big\{ P_{\rm max}(\nu)>S;M\big\} = 1 - \Bigl(1 - e^{-S}\Bigr)^M.
\label{eq:peri-pr}
\end{equation}

We use 20994 test frequencies from 0.26\,year$^{-1}$ to 2.1\,solar-day$^{-1}$, 
spaced by $10^{-4}$\,solar-day$^{-1}$. This oversamples the frequency range by 
a factor of about 2.2 and avoids underestimating the spectral power of a signal.
The number of independent frequencies is about 9500. 

\begin{figure}[!htbp]
\begin{center}
\includegraphics[width=8.8cm]{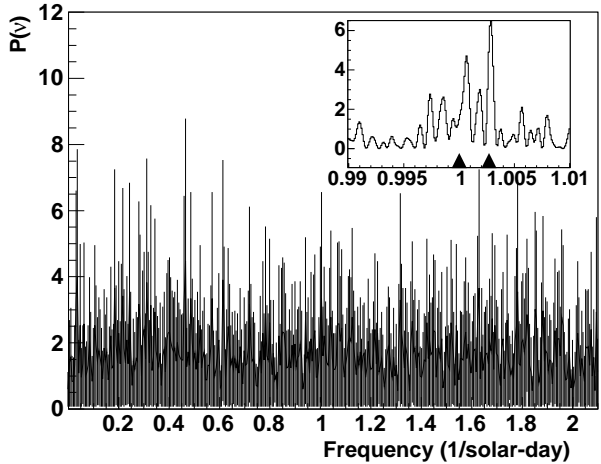}
\caption{Periodogram for opposite-sign dilepton events. 
         The solar-day and sidereal-day frequencies are indicated by the left and right 
         triangles, respectively, in the inset.}
\label{fig:periodogram}
\end{center}
\end{figure}

Figure~\ref{fig:periodogram} shows the periodogram we obtain. The largest spectral power is 
$P_{\max}(\nu) = 8.78$, for the test frequency $\nu = 0.46312$\,solar-day$^{-1}$. 
With no signal, the probability of finding a larger spectral power in our periodogram is 76\%. 
Interpolation to the sidereal frequency ($\simeq 1.00274$\,solar-day$^{-1}$) yields $P(\nu) = 5.28$, 
a value that is exceeded at 78 test frequencies. 
At the solar-day frequency, where any effects due to day/night variations in detector response 
should appear, $P(\nu) = 1.47$. 

In conclusion, we report results of a search for sidereal variations in the \CPT violation 
parameter \z that are complementary to our previous time-integrated measurements~\cite{thePRL} 
using the same events.
Neither the likelihood fits nor the periodogram method detect 
asymmetries large enough to provide evidence for \CPT and Lorentz 
violation. We have constrained the quantities $\Delta a_\mu$ of the 
Lorentz-violating standard-model extension that parameterize \CPT violation in 
\Bz -\Bzb oscillations. 

The authors are indebted to Alain Milsztajn (deceased) for his help with the 
periodogram analysis.
\input pubboard/acknow_PRL.tex

\end{document}

%% file: pubboard/authors_aug2007.tex
%
\author{B.~Aubert}
\author{M.~Bona}
\author{D.~Boutigny}
\author{Y.~Karyotakis}
\author{J.~P.~Lees}
\author{V.~Poireau}
\author{X.~Prudent}
\author{V.~Tisserand}
\author{A.~Zghiche}
\affiliation{Laboratoire de Physique des Particules, IN2P3/CNRS et Universit\'e de Savoie, F-74941 Annecy-Le-Vieux, France }
\author{J.~Garra~Tico}
\author{E.~Grauges}
\affiliation{Universitat de Barcelona, Facultat de Fisica, Departament ECM, E-08028 Barcelona, Spain }
% ...
\collaboration{The \babar\ Collaboration}
\noaffiliation

%% file: pubboard/acknow_PRL.tex
We are grateful for the excellent luminosity and machine conditions
provided by our \pep2\ colleagues, 
and for the substantial dedicated effort from
the computing organizations that support \babar.
The collaborating institutions wish to thank 
SLAC for its support and kind hospitality. 
This work is supported by
DOE
and NSF (USA),
NSERC (Canada),
CEA and
CNRS-IN2P3
(France),
BMBF and DFG
(Germany),
INFN (Italy),
FOM (The Netherlands),
NFR (Norway),
MIST (Russia),
MEC (Spain), and
STFC (United Kingdom). 
Individuals have received support from the
Marie Curie EIF (European Union) and
the A.~P.~Sloan Foundation.